\documentclass[aps,prd,floatfix,superscriptaddress,twocolumn,preprintnumbers]{revtex4} 
\usepackage{amssymb}
\usepackage{amsmath}
\usepackage{amsfonts}
\usepackage{epsfig}             
\usepackage{graphicx}
\usepackage{stackrel}
\usepackage{tabularx}
\usepackage{color}
\usepackage{dsfont}
\usepackage[T1]{fontenc}
\usepackage{natbib}
\usepackage{float}
\usepackage{placeins}

\usepackage{lipsum}

\definecolor{darkerblue}{rgb}{0.0,0.0,0.5}
\newcommand{\seq}{\begin{subequations}}
\newcommand{\sen}{\end{subequations}}

\newcommand{\eq}{\begin{eqnarray}}
\newcommand{\en}{\end{eqnarray}}

\def\nn{\nonumber}

\makeatletter
\def\@dataavailability{DATA AVAILABILITY}
\makeatother
\begin{document}
	
\title{Electron-muon conversion in nuclei and rare decays induced by LFV dark photon} 

	\author{Alexey~S.~Zhevlakov \footnote{{\bf e-mail}: zhevlakov@theor.jinr.ru}} 
        \affiliation{Bogoliubov Laboratory of Theoretical Physics, JINR, 141980 Dubna, Russia} 
        \affiliation{Matrosov Institute for System Dynamics and 
	Control Theory SB RAS, \\  Lermontov str., 134, 664033, Irkutsk, Russia } 
		
        \author{Sergey Kuleshov  \footnote{{\bf e-mail}: sergey.kuleshov@unab.cl}}
        \affiliation{Millennium Institute for Subatomic Physics at
        the High-Energy Frontier (SAPHIR) of ANID, \\
        Fern\'andez Concha 700, Santiago, Chile}
        \affiliation{Center for Theoretical and Experimental Particle Physics,
        Facultad de Ciencias Exactas, Universidad Andres Bello,
        Fernandez Concha 700, Santiago, Chile}

        \author{Valery~E.~Lyubovitskij
        \footnote{{\bf e-mail}: valeri.lyubovitskij@uni-tuebingen.de }} 
	\affiliation{Institut f\"ur Theoretische Physik, Universit\"at T\"ubingen, \\
        Kepler Center for Astro and Particle Physics, \\ 
        Auf der Morgenstelle 14, D-72076 T\"ubingen, Germany} 
	\affiliation{Millennium Institute for Subatomic Physics at
	the High-Energy Frontier (SAPHIR) of ANID, \\
        Fern\'andez Concha 700, Santiago, Chile}
	
	\author{Evgenie O.~Oleynik}
	\affiliation{Bogoliubov Laboratory of Theoretical Physics, JINR, 141980 Dubna, Russia} 
	
	\date{\today}
	
	\begin{abstract}

          We study lepton-flavor violation (LFV) effects occurring in the $e-\mu$ conversion
          in nuclei and in the rare radiative decays $\eta(\eta') \to \gamma \mu e$ of
          the $\eta$ and $\eta'$ mesons with a special impact of the sub-GeV mass vector mediator
          (dark photon). In the case of the lepton conversion, we make estimates for running and
          future experiments operating with electron beams at fixed target. Obtained results are
          implemented in analysis of the LFV decays $\eta(\eta') \to \gamma \mu e$.  
         \end{abstract}

	 \maketitle
	
\section{Introduction}

\vspace{-0.3 cm}
Study of lepton flavor violation (LFV)
is one of the promising directions for searching effects beyond the Standard Model (SM).  
During the last decades, certain experimental and theoretical progress has been achieved
in understanding and deeper studying of LFV physics: neutrino oscillations,
lepton conversion in nuclei, rare decays, etc.
In particular, the most stringent upper limit on the branching ratio
of the $\mu \to e \gamma$ decay comes from data collected at PSI (Switzerland) 
by the MEG and MEG II Collaborations: 
$\mathrm{Br}(\mu \to e \gamma) < 3.1 \times 10^{-13}$~\cite{MEGII:2023ltw}.
Before 2026, the MEG II experiment is designed to reach a sensitivity of
$\mathrm{Br}(\mu \to e \gamma) < 6 \times 10^{-14}$.  
The process $\mu \to 3e$ was studied by the SINDRUM (PSI) Collaboration giving 
the upper limit for the branching ratio
$\mathrm{Br}(\mu \to 3e) < 1 \times 10^{-12}$~\cite{SINDRUM:1987nra}.
More precise measurement of the $\mu \to 3e$ decay at PSI was proposed
by the Mu3e Collaboration, which aims for the ultimate sensitivity of
$\mathrm{Br}(\mu \to 3e) < 10^{-16}$~\cite{COMET:2025sdw,COMET:2018auw}.

Lepton conversion in nuclei is another very important tool for searching for LFV effects
with high accuracy. There are two main regimes for studying lepton conversion: lepton
scattering off nuclei at small and large energies. Coherent $\mu^- - e$ conversion
in nuclei or in muonic atoms at small energies has been studied
by the SINDRUM II Collaboration~\cite{SINDRUMII:2006dvw} at PSI. 
Reference~\cite{SINDRUMII:2006dvw} set the current measured upper limits for the ratio
\eq
R_{\mu e} = \frac{\Gamma[\mu^- + (A,Z) \to e^- + (A,Z)]}
                {\Gamma[\mu^- + (A,Z) \to \nu_\mu + (A,Z-1)]}
    \,,
    \\ \nn
    \en
which is $R_{\mu e}^{\rm Au} < 7 \times 10^{-13}$ at 90\% confidence level for
the gold stopping target. Planned experiment MECO~\cite{Popp:2000ahw} at BNL (U.S.) 
is expected to improve the SINDRUM II upper limit by at least 3--4 orders of magnitude for
$R_{\mu e}^{\rm Al}  < 5 \times 10^{-17}$ at 90\% confidence for the aluminum stopping target.
The goal of two other experiments being preparing by the COMET Collaboration
at J-PARC (Japan)~\cite{COMET:2025sdw} and the Mu2e Collaboration
at FermiLab (U.S.)~\cite{Bernstein:2019fyh} is to reach the accuracy
of $R_{\mu e}^{\rm Al} < 3 \times 10^{-17}$ at 90\% confidence for the aluminum stopping target. 
Deep inelastic lepton conversion $e + p\rightarrow \tau + X$ was searched for by 
the ZEUS Collaboration at HERA (DESY)~\cite{Chekanov:2002xz}. In the future, it is planned to
study high-energy lepton conversion at CERN SPS~\cite{Gninenko:2018num},
EIC, FCC, and LHeC colliders~\cite{Banerjee:2022xuw}. 

From the theoretical side, different models have been tested in the context of future experimental
studies of LFV effects in the NA64 experiment
at CERN SPS~\cite{Gninenko:2018num,Batell:2024cdl}, FASER$\nu$2 detector at the envisioned
Forward Physics Facility (FPF)~\cite{Feng:2022inv},
and at future colliders (EIC, FCC, LHeC)~\cite{Banerjee:2022xuw}.
There exists in literature a wide class of models and hypotheses aiming to
describe LFV phenomena, e.g., such as high-dimensional
operators~\cite{Gninenko:2018num,Dib:2018rpy},
scalar (pseudoscalar)~\cite{Dib:2018rpy,Faessler:2005hx,Gonzalez:2013rea},
ALP~\cite{Batell:2024cdl,Davidson:1981zd},
vector mediators~\cite{Dib:2018rpy,Gonzalez:2013rea,Faessler:2004ea,Heeck:2016xkh,Kachanovich:2021eqa,Araki:2022xqp,Zhevlakov:2023jzt}, etc. 
Part of research works proposed use the NA64 experiment to study LFV physics, in particular, 
lepton conversion~\cite{Gninenko:2018num,Batell:2024cdl,Zhevlakov:2023jzt}. 
These studies are possible after LS3 in the NA64 experiment~\cite{Crivelli:2023pxa, NA64:2025ddk}.

In the present paper, we focus on a role of the vector mediator (dark photon) in description
of LFV effects. In particular, we have the following main objectives:
(1) We derive the bounds on the diagonal and nondiagonal
couplings governed the LFV $\mu - e$ transition considering different rare processes.
We note that in Refs.~\cite{Zhevlakov:2023jzt,Araki:2021vhy}
parameter spaces of dark photon mediator producing LFV effects were studied in processes
with a possible search for radiation of this dark state.
(2) We concentrate on the study of fixed-target experiments using electron beams
and allowing production of a dark photon with mass in sub-GeV region. 
(3) We implement bounds for rare LFV light pseudoscalar meson decays from the dark photon model.

The study of light pseudoscalar meson decays is a one of the important tools for searching dark matter 
and studying rare physics, including $C$ and $CP$ violation~\cite{Shi:2024yfa} and LFV~\cite{Chen:2024wad}. 
In particular, decays of light pseudoscalar mesons are used to search for invisible~\cite{Gninenko:2023rbf,NA64:2024mah} 
and visible~\cite{COHERENT:2022pli} dark photon modes. A broad scientific program, including LFV decays and 
rare decays of the $\eta$ and $\eta^\prime$ mesons, is indicated in the proposals of the REDTOP~\cite{REDTOP:2022slw} 
and eta-HIAF~\cite{Chen:2024wad} factories. The eta-HIAF factory is expected to yield approximately  $10^{12} \div 10^{13}$ $\eta$ mesons 
during one year~\cite{Mushtaq:2025qfh,Chen:2024wad} that will be a good test of rare decay physics. In according to interest for study 
LFV $\eta$ and $\eta^\prime$ meson decays at HIAF factory we made the estimates based on current limits for vector portal.  
We shown that such rare decays can be more sensitive to LFV effects than lepton conversion at GeV energies of electron beams.  

The paper is organized as follows. In Sec.~\ref{Sec:framework},we present details of
inclusion of dark photons in our quantum field formalism. 
In Sec.~\ref{Sec:LFVphysics}, we discuss kinematical and dynamical aspects of
the LFV physical processes, which are studied in our paper.
In particular, we focus on two- and three-body rare decays of muon and $e-\mu$ lepton
conversion in nuclei using electron beams at fixed-target experiments.
Later, in Sec.~\ref{Sec:bouns}, we briefly discuss what follows from results for
the LFV rare decays of the $\eta$ and $\eta^\prime$ mesons. In Sec.~\ref{Sec:summary},
we summarize our results and findings. Additionally, some complementary formulas are included in the Appendix~\ref{Sec:AppA}.

\section{Inclusion of dark photons in Quantum Field Formalism}
\label{Sec:framework}
  
In our paper, we follow quantum field formalism allowing one to include the dark photon
developed in Ref.~\cite{Kachanovich:2021eqa}. In particular, the phenomenological
Lagrangian involves singlet scalar field $\sigma$, dark photon $A'$, and Dirac dark fermion
$\chi$. The mass of the dark photon $m_{A'}$ is generated via the Stueckelberg
mechanism (Stueckelberg portal)~\cite{Stueckelberg:1938zz} by extension
of the derivative acting on the scalar field $\sigma$ to covariant one containing mass
of the $A'$. The corresponding model Lagrangian reads~\cite{Kachanovich:2021eqa}
\begin{eqnarray}\label{eq:DS-Lagr-gf}
{\cal L}'_{\rm DS} &=&  - 
\frac{1}{4} \, {A}_{\mu\nu}' {A}^{\prime \mu\nu} 
\,+\, \frac{m_{A'}^{2}}{2} A'_{\mu} A^{\prime \mu} \nonumber\\
\,&+&\, \bar\chi \, (i\not\!\!D_{\chi} - m_\chi) \, \chi   
-  \frac{1}{2\xi}\left(\partial_{\mu}A^{\prime \mu} \right)^{2}\\
&+& \frac{1}{2} \partial_{\mu}\sigma\partial^{\mu}\sigma - \xi \frac{m_{A'}^{2}}{2}
\sigma^{2} \,, \nonumber
\end{eqnarray} 
where $\xi$ is the gauge-fixing parameter.
Note, the interaction of the dark photon with charged SM fermions 
can include both diagonal and nondiagonal couplings	
\begin{eqnarray}
\label{eq:A'-psi-1} 
{\cal L}_{\rm A'\psi} &=& \epsilon e A'_\mu(x) \, 
\sum\limits_{i,j=e,\mu,\tau} \, \bar\psi_i(x) \, \gamma^\mu \, G_{ij} \, \psi_j(x) \,, 
\nonumber\\
 G_{ij} &=& - \delta_{ij} + g^{V}_{ij} + g^{A}_{ij} \, \gamma_5  \,,
\end{eqnarray}
\vspace{0 cm}
where $g_{ij}^V$ and $g_{ik}^A$ are the vector and axial-vector dimensionless couplings,
respectively; $\epsilon$ is the QED-dark-photon mixing parameter. The first term comes from general 
kinetic mixing of the photon and dark photon after shift of photon field~\cite{Pospelov:2008zw}.  
Coupling proportional to $e \epsilon g^V_{ll}$ can be induced by direct interaction of the dark photon 
with matter based on a different mechanism. 
For example, such scenarios have been proposed  in Refs.~\cite{Kachanovich:2021eqa, Belyaev:2022shr}. 
In the following, we use the following notations $g_{ll} =e\epsilon(-\delta_{ii} +g^V_{ii})$ and $g_{lf}=e\epsilon g_{ij}^{A,V}$
for diagonal and nondiagonal couplings, respectively, that is needed for analysis. The $\delta_{ij}$ is delta Kronecker symbol.  
In our applications we focus on the use of the vector couplings. 

In the framework of this work, we consider different scenerios of interaction dark photon with SM matter, for diagonal couplings:
\\
\begin{itemize}
	\item \textit{The General dark photon portal:}
	\\ This portal is connected only with general kinetic mixing 
	term~\cite{Holdom:1985ag} $\mathcal{L}=\dfrac{\epsilon}{2}F_{\mu\nu} F^{\prime \mu\nu}$ 
	where $\epsilon$ is the kinetic mixing parameter between the two
	$U(1)$ gauge symmetries, and $F_{\mu\nu}$ and $F^{\prime \mu\nu}$ are the strength
	tensors of SM electromagnetic and DM dark gauge (dark photon) fields, respectively. 
	By shift of the electromagnetic (EM) field $A_\mu \to A_\mu-\epsilon A'_\mu$ we can obtain interaction  dark photon 
	with charge current of lepton or quark fields.  In results  $g_{ll} =e\epsilon(-\delta_{ii} +g^V_{ii})$ 
	includes interaction with all fermions, in general.  
	
\item \textit{The non-minimal leptophilic dark photon portal:}
In this scenario, the dark photon couples directly only to leptons, with no tree-level kinetic mixing. 
The direct coupling is parametrized as \(g_{ll} = e\epsilon (g^V_{ll})\). Interactions with quarks are 
not present at tree level, but are generated radiatively through fermion loops involving leptons and 
a virtual photon. This loop-induced coupling is suppressed by a factor of \(\sim e^2/(16\pi^2)\) and 
exhibits a logarithmic dependence on the lepton mass scale, \(\ln(m_l^2/\mu^2)\).

\item \textit{The non-minimal hadrophilic dark photon portal:}
This is the complementary case, where the dark photon couples directly only to quarks, 
gain with no tree-level kinetic mixing. The direct coupling is \(g_{qq} = e\epsilon (g^V_{qq})\). 
Interactions with leptons are generated indirectly through quark-loop diagrams and are suppressed 
by \(\sim e^2/(16\pi^2)\), with a logarithmic dependence on the quark mass scale, \(\ln(m_q^2/\mu^2)\).
\end{itemize}	
\newpage
\vspace{1 cm}
\onecolumngrid
\begingroup
\begin{figure*}[t]
	\includegraphics[width=0.95\textwidth, trim={3cm 13.6cm 6cm 13.5cm},clip]{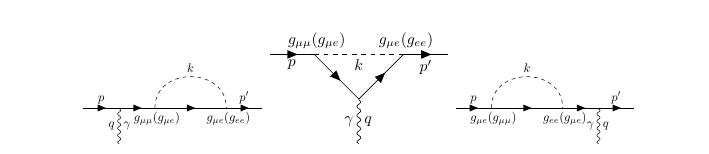}
	\caption{
		Gauge-invariant set of Feynman diagrams induced by
		dark photon mediator and contributing to the $\mu \to e \gamma$ decay}
	\label{DiagG2}
\end{figure*}
\endgroup	
\twocolumngrid

A different scenario will give a different dependence in interaction, including vector meson and dark photon 
mixing which is considered here at estimation LFV decays of $\eta$ and $\eta'$ mesons.  
In this time exist  upper experimental limits on different scenerios which will be used here. 
 	
\section{LFV decays}
\label{Sec:LFVphysics} 

In this section, we briefly discuss kinematical and dynamical aspects
of the following LFV processes: decays of muons $\mu \to e \gamma$
and $\mu\to 3e$ and $e-\mu$ conversion in nuclei. Later, we will see 
that rare decays give strong bounds for parameter space of models.

\subsection{Decay $\mu \to e \gamma$}

A gauge-invariant set of Feynman diagrams induced by the
dark photon mediator and contributing to the
$\mu \to e \gamma$ decay is shown in Fig.~\ref{DiagG2}. 
Corresponding manifestly gauge-invariant leptonic current
in momentum space has the form 
\eq
J^\mu(p,p') &=& \bar{u}_e(p') \Gamma^\mu(p,p') u_\mu(p)
\,,\nonumber\\
\Gamma^\mu(p,p') &=& 
\gamma^\mu_\perp  F^{e\mu}_1(q^2)
+ \frac{i\sigma^{\mu \nu}q_\nu}{m_\mu + m_e} F^{e\mu}_2(q^2)\,,
\en
where $\gamma^\mu_\perp = \gamma^\mu - \dfrac{q^\mu \not\! q}{q^2}$ is the
Dirac matrix orthogonal to the photon momentum $q=p-p'$ with
$q_\mu \, \gamma^\mu_\perp = 0$.

Expression for the form factors $F^{e\mu}_i(q^2)$ is given in the Appendix.
In the limit $m_\mu=m_e$ and $m_{A'}\to 0$, the $F^{e\mu}_2(0)$ is equal to
$\dfrac{g_{ll}g_{lf}}{4\pi^2}$, which is the LFV analog of
the Schwinger contribution $\dfrac{\alpha}{2\pi}$
to the anomalous magnetic moment of the charged fermion with spin $1/2$ in QED. 

The decay width of the $\mu \to e \gamma$ transition reads 
\eq
\Gamma (\mu \to e \gamma)
= (g_{\mu e}g_{ll})^2\frac{\alpha}{2} \, m_\mu \, \Big(1 - \frac{m_e}{m_\mu}\Big)^2 \,
\Big[F^{e\mu}_2(0)\Big]^2 \,. 
\en
Note, the Dirac form factor $F^{e\mu}_1(q^2)$ is finite for $q^2 \neq 0$,
while it has divergence at $q^2 = 0$. This divergence is removed
by introducing an appropriate counterterm in analogy with QED.
Therefore, the renormalized Dirac form factor has the form
$F^{e\mu}_{1; r}(q^2) = F^{e\mu}_1(q^2)-F^{e\mu}_1(0)$. It means that
$F^{e\mu}_{1; r}(0) = 0$; i.e., LFV is absent for real photon emission. 
The expression for the $F^{e\mu}_{1; r}(q^2)$ is presented in the Appendix~\ref{Sec:AppA}.
In the limit $m_e=m_\mu$ and $m_{A'}=0$, we reproduce the well-known QED result
for the one-loop vertex correction~\cite{Peskin:1995ev}. 

\begin{figure}[t]
	\includegraphics[width=0.25\textwidth, trim={0cm 0cm 0cm 0cm},clip]{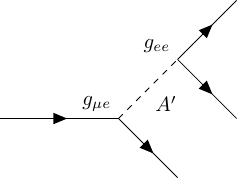}
	\caption{Diagram contributing to the $\mu \to 3e$ decay 
          generated by the dark photon mediator
          due to its nondiagonal coupling with SM leptons.} 
	\label{DiagG3}
\end{figure}

\subsection{Decay $\mu \to 3e$}

The diagram contributing to the $\mu \to 3e$ decay, which is induced
by the nondiagonal coupling of the dark photon with SM leptons, is
shown in Fig.~\ref{DiagG3}. 
The $\mu \to 3e$ decay width in the limit of massless electron
$m_e = 0$ is given by 
\eq
\Gamma(\mu \to 3e) = \frac{(g_{ll}g_{\mu e})^2}{256 \pi^3 m_\mu^3}
\, \int\limits_0^{m_\mu^2} ds_2\!\!\!\!\!
 \int\limits_0^{m_\mu^2-s_2}\!\!\! ds_1\,\,
\frac{1}{2}\sum_{\rm pol} |M_{\rm inv}|^2\,,\,\,
\en
where $ \frac{1}{2} \, \sum_{\rm pol} |M_{\rm inv}|^2$ is the square of
the invariant matrix element of the $\mu \to 3e$ decay
averaged over the muon spin and polarizations
of fermions, 

\onecolumngrid
\onecolumngrid
\begin{center}
	\begin{table}[t!]
		\centering
		\caption{Parameters of the fixed-target experiments NA64$_e$,
			LDMX, BDX, and darkSHINE: target parameters ($A, Z$),
			energy of scattering beam $E$, target density $\rho_T$, 
			first radiation length $X_0$, effective thickness of
			the target ($L_T$),
			and planned accumulate of electrons on target (EoT).
		}
		\begin{tabular}[t]{rccccccc}
			\hline
			\hline
			& $e$-conv. &$A$\, ($Z$) & $E$ (GeV)& $\rho_T$ (g cm$^{-3}$)
			& $X_0$ (cm)& $L_T$ (cm) & Projected EoT\\
			\hline
			NA64$_e$:& $e N\to   N \mu$&207 (82)&100 & 11.34 & 0.56  &0.5 &
			$5 \times 10^{12}$\\
			LDMX: & $e N\to   N  \mu$ & 27 (13) & 16&2.7&8.9&3.56 & $10^{18}$ \\
			BDX: & $e N\to   N  \mu$  & 27 (13) & 11&2.7&8.9&3.56 & $10^{22}$ \\
			darkSHINE: & $e N \to N \mu $ & 184 (74)&8&19.3&0.35&0.35 & $10^{16}$ \\
			\hline
			\hline
			\label{ParamTable}
		\end{tabular}
	\end{table}%
\end{center}
\twocolumngrid
\twocolumngrid

\eq
\frac{1}{2} \, \sum_{\rm pol} |M_{\rm inv}|^2 \!\!
&=&\!\!
8 \bigg(
\frac{(s_1 + s_2) (m_\mu^2 - s_1 - s_2) + s_2 (m_\mu^2 - s_2)}
     {2(s_1-m_{A'}^2)^2} \nn \\
&+&
\frac{(s_1 + s_2) (m_\mu^2 - s_1 - s_2) + s_1 (m_\mu^2 - s_1)}
     {2(s_2-m_{A'}^2)^2}  \nn	\\
&-&
     \frac{(s_1 + s_2) (m_\mu^2 - s_1 - s_2)}{(s_1-m_{A'}^2)(s_2-m_{A'}^2)}\Bigg)\,.
\en
Here we use the Mandelstam variables,  
\begin{align}
	s_1 = (p_1 + p_2)^2 = (p - p_3)^2 \,,\nn\\
        s_2 = (p_2 + p_3)^2 = (p - p_1)^2 \,,\nn\\
	s_3 = (p_1 + p_3)^2 = (p - p_2)^2\,, \nn\\
	s_1 + s_2  + s_3 = s = m_\mu^2 \,.
\end{align}	

\subsection{Lepton conversion in fixed-target experiments}
\label{Sec:conv}

In this section, we focus on the process of lepton $e-\mu$ conversion. Here our main
purpose is to understand possible opportunities for searching this process in fixed-target experiments. In particular, we strict to experiments, which are able to produce
lepton beams with huge intensity and aim to detect feebly interactive particles of
dark matter or mediators between SM and DM. One can expect that in fixed-target
experiments $e-\mu$ conversion should have a good signal, because a produced muon will
have an energy close to beam energy, and therefore it can be registered.
Production of such an energetic muon will be the tag for the $e-\mu$ conversion. 
Note, the use of muon beam in fixed-target experiments for searching $\mu - e$ conversion
is more problematic due to small radiation and dissipation lengths of electron in a target.
The $\mu - \tau$ and $e - \tau$ conversions have  difficulties in isolating the signal
from $\tau$ lepton decays.

One of the best candidates to study $e-\mu$ conversion in the regime of fixed target by
using electron beam is the NA64$_e$ experiment at CERN SPS~\cite{Andreas:2013lya,NA64:2023wbi}. 
At present, the NA64$_e$ setup is able to collect $\sim 10^{12}$ electrons on target. 
Additionaly, in the future it is planned to search for dark matter using 
electron beams with different types of targets in the experiments LDMX
at SLAC (U.S.)~\cite{Berlin:2018bsc,LDMX:2018cma,Ankowski:2019mfd},
BDX at FermiLab (U.S.)~\cite{BDX:2016akw}, and
DarkSHINE at SHINE (China)~\cite{darkSHINE:2024guq}.
The experiments NA64$_e$, LDMX, and DarkSHINE are based on the use of missing energy
and momentum techniques. The BDX experiment for searching dark matter is based on beam dump
with a possibility to observe visible decay after the wall~\cite{BDX:2016akw}.
All experiments have a calorimeter system and track system, which can register energetic
muons inside. We collect the main parameters of the fixed-target experiments NA64$_e$, LDMX,
BDX, and DarkSHINE in Table~\ref{ParamTable}. 

The number of produced muons in the reaction of the $e - \mu$ conversion is equal
\eq
N_{\mu} \simeq \mbox{EoT}\cdot \frac{\rho_T N_A}{A} L_T \int\limits
dx \frac{d \sigma(E)_{e \to \mu}}{dx} \,.
\en
where $N_A$ is the Avogadro number, and $d\sigma_{e \to \mu}/dx$ is the  differential cross section
of the lepton conversion $e N_Z \to \mu N_Z$ on nuclear $N_Z$ with atomic number $A$
and charge $Z$. Other parameters are collected in Table~\ref{ParamTable}.

Kinematics of lepton conversion is specified by the following set of
the Mandelstam variables: 
\begin{align}
s = (k_1 + p_1)^2 = (k_2+p_2)^2   \,, \nn\\
t = (k_2 - k_1)^2 = (p_2 - p_1)^2  \,, \nn\\
u = (k_1 - p_2)^2 = (k_2 - p_1)^2  \,, \nn\\
s + t + u = 2M_{N_Z}^2 + m_\mu^2 + m_e^2 \,, 
\end{align}
\noindent where $p_1$ and $p_2$, $k_1$ and $k_2$ are
the initial and final nuclear momenta and
the electron and muon momenta, respectively; 
$M_{N_Z}$ is the nuclear mass.

In general, the interaction of an electromagnetic field $A_\mu$ and
nuclear can be effectively represented as~\cite{Schwartz_book}:
\eq
\mathcal{L}^{\mathrm{eff}}_{N_Z} = - eA_\nu  J_{N_Z}^\nu \,, 
\en
\noindent where $J_{N_Z}^\nu$ is hadronic current of interaction nuclear
with EM field $A_\nu$. For heavy nucleus, one can exploit a spin-0 form factor
with a good accuracy~\cite{Beranek:2013yqa}.  
Thus, the hadronic current in the momentum space reads
as~\cite{Beranek:2013yqa,Perdrisat:2006hj}:
\eq
J_{N_Z}^\nu=F_{N_Z}(t) \, (p_1+p_2)^\nu \,.
\en

\begin{figure}[t!]
	\includegraphics[width=0.22\textwidth, trim={0cm 0cm 0cm 0cm},clip]{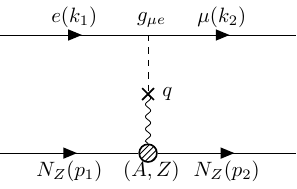}
	\quad
	\includegraphics[width=0.22\textwidth, trim={0cm 0cm 0cm 0cm},clip]{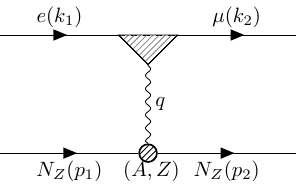}
	\caption{Diagrams contributed to the lepton $e - \mu$ conversion in nuclei due
       to LFV dark photon: diagram induced by mixing of QED and dark photon (left); 
       (b) diagram induced by form factor relevant for the $\mu \to e \gamma$ transition
       (right). Dashed triangle blob in right diagram represents the vertex
       function $\Gamma_\nu(k_1,k_2)$.}
	\label{DiagConv}
\end{figure}

Lepton conversion in nuclei is contributed by two diagrams
shown in Fig.~\ref{DiagConv}: (left panel) diagram induced by mixing of QED
and dark photon (right panel) diagram induced by form factor relevant for the $\mu \to e \gamma$
transition. 
Dashed triangle blob in right diagram represents the vertex
function $\Gamma_\nu(k_1,k_2)$.
Matrix element of lepton conversion is  
given by the sum of two terms $M^{\rm conv}_1$ and $M^{\rm conv}_2$
corresponding to diagrams shown in Fig.~\ref{DiagConv}:
\eq
M^{\rm conv} = M^{\rm conv}_1 + M^{\rm conv}_2 \,, 
\en
where 
\eq
M^{\rm conv}_1 &=& \frac{e^2 g_{ll}g_{e\mu}}{q^2 - m_{A'}^2}
\, \bar{u}_e(k_2) \gamma_\nu u_\mu(k_1) \, J_{N_Z}^\nu\,,\\
M^{\rm conv}_2 &=& \frac{e^2 g_{ll}g_{e\mu}}{q^2 - m_{A'}^2}
\,  \bar{u}_e(k_2) \Gamma_\mu(k_1,k_2) u_\mu(k_1)
\, J_{N_Z}^\nu  \,. \quad
\en
In $M^{\rm conv}_1$ the mixing is induced by lepton loop between dark photon and external photon 
coming from nuclei. Here it is proposed that direct interaction of dark photon with baryons does not exist 
and only occurs via mixing generated by the lepton loop. 
Leprton-loop contribution is collected in the product of two coupling $e g_{ll}$. 
Differential cross section is given by 
\eq
\frac{d\sigma}{dt} = \cfrac{\frac{1}{2}\sum\limits_{\rm pol} \,
|M_{conv}|^2}{16\pi \, (s - M_{N_Z}^2)}
\,.
\en
The mixing parameter $\epsilon$ is considered as a free parameter.
The interaction with nuclear is parametrized by taking into account a general
electric form factor which includes $Z^2$ and $Z$ terms describing 
"\textit{elastic}" and "\textit{inelastic}" effects, respectively.
The form of this nuclear form factor is parametrized by Tsai~\cite{Tsai:1973py,Kim:1973he}
and Bjorken \textit{et al.}~\cite{Bjorken:2009mm,Bjorken:1988as} as
\eq
F_{N_Z}^2(t) &=& \biggl(\frac{a^2 t}{1+a^2 t}\biggl)^2
\biggl(\frac{1}{1+\dfrac{t}{d}}\biggl)^2 Z^2 \nn\\
&+& \biggl (\frac{a^{\prime 2} t}{1+ a^{\prime 2}t} \biggl)^2
\left(\frac{1+\dfrac{t}{4 m_p^2}(\mu_p^2 -1)}{(1+\dfrac{t}{0.71 GeV^2})^4}\right)^2
Z\, , \qquad 
\en 
where the parameters of nuclear form factor are specified:
$a=111 \, Z^{-1/3}/m_e$ and $d=0.164 \, A^{-2/3}$ GeV$^2$ are 
the screening and nucleus size parameters, respectively~\cite{Bjorken:2009mm} and
$a^\prime= 773 \, Z^{-2/3}/m_e$, $\mu_p = 2.793$ (proton magnetic
moment)~\cite{ParticleDataGroup:2024cfk} are parameters for inelastic scattering.

\section{Results}
\label{Sec:bouns}

\begin{figure}[t]
	\centering
	\includegraphics[width=0.99\linewidth, trim={0cm 0cm 0cm 0cm},clip]{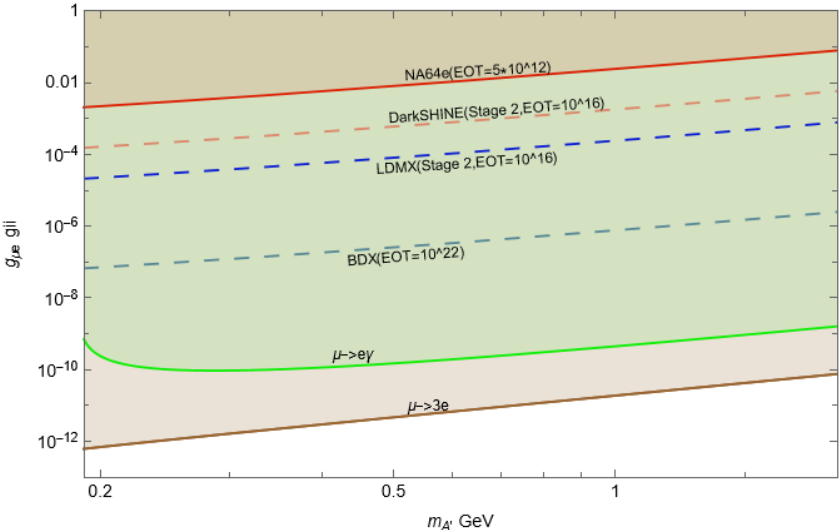}
	\caption{$m_{A'}$ dependence of bounds for the products of diagonal and nondiagonal
       couplings of dark photon with SM fermions. 
       Shaded region corresponds to available data from limits on rare muon LFV decays and
       for current statistics of the NA64 experiment. Dashed lines correspond to estimates of
       future beam dump experiments with electron beams.
       Parameters used here are presented in Table~\ref{ParamTable}.}
	\label{resPlot}
\end{figure}

In this section, we present our results for the bounds on the LFV couplings
extracted from analysis of rare LFV decays. We hope that our predictions
will be useful for running and future experiments with 
huge intensity electron beams in fixed-target experiments. Unfortunately,
the bounds on the LFV couplings derived from study of the $e - \mu$ conversion in nuclei
using electron beam with exchange by vector boson between LFV lepton current and
nuclear current are less stringent than existing bounds from LFV muon decays.
Obtained bounds for diagonal and nondiagonal couplings as functions of mass of
dark photon $m_{A'}$ are shown in Fig.~\ref{resPlot}. It seems that currently
in fixed-target experiments it is impossible to get the level of sensitivity achieved
by decay studies. 

\subsection{Bounds for LFV couplings from rare $\eta$ and $\eta^\prime$ meson decays}
\label{Sec:LFVETAS}

Using obtained results for combination of diagonal and nondiagonal coupling,
we have a possibility to estimate branchings of the LFV decays of $\eta$ and $\eta^\prime$ mesons.
An experimental study of these decays is planned in the $\eta$ and $\eta^\prime$ factory REDTOP
at FermiLab (USA)~\cite{REDTOP:2022slw}. The vector mediator is a simple model, which can give
bounds for the three-body LFV decay widths of $\eta(\eta^\prime) \to \gamma \mu e$
using existing limits from LFV decay and lepton conversion.
  \newpage
\onecolumngrid
\onecolumngrid
\begin{figure}[t!]
	\includegraphics[width=0.22\textwidth, trim={0cm 0cm 0cm 0cm},clip]{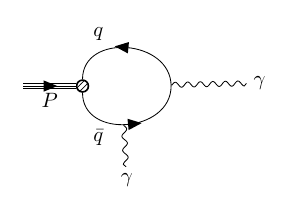}
	\quad
	\includegraphics[width=0.22\textwidth, trim={0cm 0cm 0cm 0cm},clip]{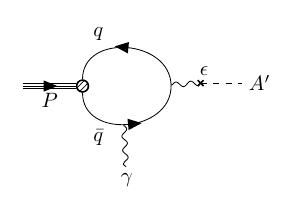}
	\quad
	\includegraphics[width=0.22\textwidth, trim={0cm 0cm 0cm 0cm},clip]{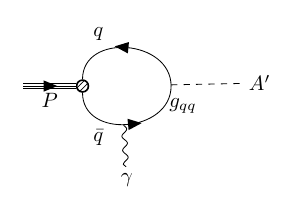}
	\quad
	\includegraphics[width=0.26\textwidth, trim={0cm 0cm 0cm 0cm},clip]{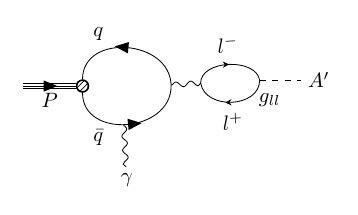}
	\\
	a) \qquad \qquad\qquad\qquad\qquad \qquad\qquad b)\qquad \qquad\qquad\qquad \qquad\qquad c) \qquad\qquad\qquad\qquad\qquad\qquad d) 
	\caption{Feynman diagrams that show vertices of decay pseudoscalar meson into two photons (a) 
		or one photon and a dark photon in different scenarios: (b) general, (c) hadrophilic, and (d) leptophilic.}
	\label{DiagPGA}
\end{figure}
\twocolumngrid
 In general, such bound should be
weakly changed in dependence
of type of mediator. For other LFV decays, such as two-body decays $\eta\to e\mu$ and
$\eta^\prime\to e\mu$, strict limits from conversion were obtained in
Ref.~\cite{Hoferichter:2022mna}.

The LFV decays $\eta(\eta^\prime) \to \gamma \mu e$ are related to the
Dalitz lepton-flavor-conversing decays $\eta(\eta^\prime) \to \gamma l^+l^-$
because matrix elements of both
types of processes contain anomalous pseudoscalar-vector-vector couplings:
$P \gamma A'$ and  $P \gamma \gamma$ (the Wess-Zumino-Witten chiral anomaly), respectively.
A nice point is that we can relate $P \gamma A'$ and $P \gamma \gamma$ couplings
using the QED-dark-photon mixing. In particular, 
starting with phenomenological Lagrangian governing
 the chiral anomaly

\eq
{\cal L}_{P\gamma\gamma} &=&
\frac{e^2}{8} \, g_{P\gamma\gamma} \,  
\epsilon_{\mu\nu\alpha\beta} \, P(x) \,  
F^{\mu\nu}(x) \, F^{\alpha\beta}(x) \,,
\en
where $F^{\mu\nu}$ is the stress tensor of electromagnetic field,
$\epsilon_{\mu\nu\alpha\beta}$ is the antisymmetric Levi-Civita tensor, 
and $g_{P\gamma\gamma}$ is the $P\gamma\gamma$ coupling. 
This vertex in shown in Fig.~\ref{DiagPGA}(a).
We derive phenomenological Lagrangian describing the coupling 
of pseudoscalar meson $P$ with QED and dark photons,
\eq
\label{LPgammaAprime}
{\cal L}_{P\gamma A'} &=&
\frac{e^2}{8} \, 
g_{P\gamma A'} \,  
\epsilon_{\mu\nu\alpha\beta} \, P(x) \,  
F^{\mu\nu}(x) \, F^{\prime \alpha\beta}(x) \,, 
\en
where $g_{P\gamma A'}$ is the $P\gamma A'$ coupling, which for three different scenarios 
[corresponding Feynman diagrams are shown 
in Figs.~\ref{DiagPGA}(b)-~\ref{DiagPGA}(d)] are fixed as:
\begin{itemize}
	\item[(1)]  General scenario: $g_{P\gamma A'} = \epsilon g_{P\gamma\gamma}$. In this scenario we base on the kinetic mixing of 
	the QED and dark photons~\cite{Holdom:1985ag}, which after the shift of one of the EM field $A_\mu \to A_\mu-\epsilon A'_{\mu}$ leads 
	to the coupling of the pseudoscalar meson pair of QED and dark photons.  
\item[(2)] Hadrophilic scenario: $g_{P\gamma A'} = (g_{qq}/e) g_{P\gamma\gamma}$~\cite{Escribano:2022njt,Tulin:2014tya}. 
In this scenario, the dark photon directly interacts with quarks, 
\item[(3)] Leptophilic scenario: $g_{P\gamma A'} = \cfrac{e g_{ll}}{16\pi^2} 
\ln(m_l^2/\mu^2) g_{P\gamma\gamma}$~\cite{Gherghetta:2019coi}. 
In this scenario dark photon directly interacts with leptons, and therefore its coupling with the QED photon occurs 
via the lepton loop, where $\mu$ is a renormalization scale 
and $m_l$ is the mass of lepton propagating in the loop.
\end{itemize}
Note, the decay width of the anomaly process $P \to \gamma \gamma$ 
in terms of coupling $g_{P\gamma\gamma}$ reads
\eq 
\Gamma(P \rightarrow \gamma\gamma)
&=& \frac{\pi \, \alpha^2}{4} \, g^2_{P\gamma\gamma}
\, m_P^3 \,,
\en
where $m_P$ is the mass of the respective pseudoscalar meson.
$P \rightarrow \gamma A'$ decay width can be written as 
\eq 
\Gamma(P \rightarrow \gamma A') =
\frac{\epsilon^2}{2} \, (1 - x_{A'P})^3
\, \Gamma(P \rightarrow \gamma \gamma) 
\,,
\en
where $x_{ij} = m_i^2/m_j^2$ is the ratio of squared masses of
corresponding particles.

With taking into account intermediate 
vector mesons $V = \rho^0,\omega,\phi$
we get for the decay width 
\eq
\label{WithRhoDecayPS}
\hspace*{-.75cm}
& &\Gamma(P \to \gamma \mu e) = \frac{\alpha g_{\mu e}^2}{1024 \pi m_P^3}
\int\limits_{m_\mu^2}^{m_\eta^2} \frac{dq^2}{q^4}
\nonumber\\
\hspace*{-.75cm}
&\times&
\frac{A(m_P^2,m_\mu^2,q^2)}
     {(m^2_{A'}-q^2)^2+\Gamma^2_{A'\to \mu e} m^2_{A'}} g_{P\gamma A'}^2
\,, 
\en
where
\eq
&&A(m_P^2,m_\mu^2,q^2)=(m_{\mu}^2-q^2) \Big((m_P^2-q^2)^3(q^2-m_\mu^2)^2\nn\\
&&-(m_P^2-q^2)^3(q^4-2m_\mu^4)-(m^6_P-q^6)m_\mu^2q^2\Big)\,.
\en
In the case with including intermediate vector meson states,
the coupling $g_{P\gamma A'}^2$ is changed as
\eq
g_{P\gamma A'}^2 \to \biggl[g_{P\gamma A'}^2
+ 2 g_{P\gamma A'} Re(g_{P\gamma A'}^R(q^2))
+ |g_{P\gamma A'}^R(q^2)|^2
\biggr] \,,\nonumber
\en
with mixing photon and vector meson  
\eq
g_{P\gamma A'}^R(q^2) = 
\sum_{V = \rho^0, \omega, \phi}  
\frac{q^2 \, g_{VP\gamma} \, g_{VA'}}{q^2 - m^2_{\rho} + i \Gamma_{\rho} m_{\rho}} \,,
\en
where coupling $g_{P\gamma A'}$ for different scenarios was specified
after Eq.~(\ref{LPgammaAprime}), 
end coupling $g_{VA'}$ defining the $A' \to V$ transition is
fixed for three scenarios as: 
\begin{itemize}
\item[(1)] $g_{VA'} = \epsilon g_{V\gamma}$ for general dark photon scenario;
\item[(2)] $g_{VA'}=g_{V\gamma}(g_{qq}/e)$ for the hadrophilic scenario \cite{Escribano:2022njt,Tulin:2014tya};
\item[(3)]\ $g_{VA'}= \cfrac{g_{ll} e}{16\pi^2} \ln(m_l^2/\mu^2) g_{V\gamma}$
for the leptophilic scenario. 
\end{itemize}
The hadrophilic  and leptophilic scenarios are specified by 
direct interaction of dark photon only with quarks or leptons, respectively. The general scenario is 
based on the general mixing mechanism allowing direct interaction of the dark photon with both quarks 
and leptons.

The $g_{\rho^0\eta\gamma}$ coupling is fixed from decay width
\eq
\Gamma(V \rightarrow P\gamma)
= \frac{\alpha}{24} \, g^2_{VP\gamma}
\, m_V^3 \, (1 - x_{PV})^3 \,.
\en 
The couplings $g_{VP\gamma}$
are fixed from data on the corresponding
$P \to V \gamma$ decay
widths
\eq
\Gamma(P \rightarrow V\gamma)
= \frac{\alpha}{8} \, g^2_{VP\gamma}
\, m_P^3 \, (1 - x_{VP})^3 \,.
\en 
Numerical values of the coupling  $g_{P\gamma\gamma}$ and $g_{P V\gamma}$ fixed from
data~\cite{ParticleDataGroup:2024cfk} are:
\eq
g_{\eta\gamma\gamma} &=& 0.274 \, {\rm GeV}^{-1}
\,, \quad g_{\eta'\gamma\gamma} = 0.341 \, {\rm GeV}^{-1} \,,
\nonumber\\
g_{\rho^0\eta\gamma}  &=& 1.55 \, {\rm GeV}^{-1}\,, \quad 
g_{\rho^0\eta'\gamma}  = 2.73 \, {\rm GeV}^{-1}\,.
\en
The  specific neutral vector meson
$g_{V\gamma}$ is equal to
$g_{\rho\gamma} = 0.202$ for $\rho^0$,
$g_{\omega\gamma} = 0.059$ for $\omega$, and
$g_{\phi\gamma} = 0.075$ for $\phi$. 

In the limit $m_e\to 0$ the decay width $\Gamma_{A'\to \mu e}$ 
is given by 
\eq
\Gamma_{A'\to \mu e} = \frac{g_{\mu e}^2}{12\pi} m_{A'}\sqrt{1-y_\mu^4}
\, \left(1-2y_\mu^2-2y_\mu^4\right),
\en 
where $y_l = m_l/m_{A'}$. The decay width of the dark photon into a lepton-antilepton pair of the same flavor is given by 
\eq 
\Gamma_{A'\to \bar{l}l } = \frac{g_{ll}^2}{12 \pi} 
\, m_{A'} \, 
\sqrt{1 - 4 y_l^2}(1 + 2 y_l^2)  \,.  
\en
where  $g_{ll}$ will be dependent on scenario.

For heavy mass of the dark photon $m_{A'} > m_\eta$, the decay width
$\Gamma(P \to \gamma \mu e)\sim m_P (g_{ll}g_{\mu e})^2$. Current bounds for production of diagonal
and nondiagonal coupling are at level $10^{-10}$;
hence $\mathrm{Br}(\eta (\eta') \to \gamma \mu e)$
will be at level $\lesssim 10^{-18}$. 
Maximal value for this decay width can be reached for the dark photon mass below $\eta$ meson
mass, i.e., in the region  $m_{A'}<m_\eta$, where resonance
production is possible. In sub-GeV mass region, the decay width for the dark photon is of order 
$\Gamma_{A'\to \mu e} \sim m_P\times O(g^2_{\mu e})$, and decay width $\Gamma(\eta (\eta') \to \gamma \mu e)$
can be computed using known narrow-width 
approximation~\cite{ParticleDataGroup:2024cfk,Chivukula:2017lyk, Uhlemann:2008pm},
\eq
&&\int\limits_{s_-}^{s_+} ds \,
\frac{f(s)}{(s-m^2)^2 + m^2 \Gamma^2} \approx
\int\limits_{s_-}^{s_+} ds \, 
\frac{\pi}{m \Gamma} \, f(s) \, \delta(s-m^2)
\nonumber\\
&& \qquad=\frac{\pi}{m \Gamma} f(m^2) 
\en 
for $\Gamma \ll m$ and  $s_- < m^2 < s_+$.  
The $\Gamma$ is decay width, which is the sum of all channels and final results will be dependent upon which mode is dominant.

For the upper limit, we used in narrow-width approximation decay width $A'\to \mu e$ if nondiagonal coupling is dominant.
 As a result,  $\Gamma(P \to \gamma \mu e)$ is proportional to $g_{ll}^2$.
In this case, the general, hadrophilic, or leptophilic scenario for diagonal interaction
couplings  could occur. Other scenarios are described in Ref.~\cite{Bauer:2018onh}. 
All scenarios have bounds from experiments that studied dark matter or can be recalculated
from one to another. For the leptophilic scenario, 
there exist bounds from invisible mode from the BABAR~\cite{BaBar:2017tiz} and
NA64~\cite{NA64:2023wbi} experiments. For the hadrophilic scenario, current limits for $m_{A'}<m_\eta$
are established from the COHERENT~\cite{COHERENT:2022pli} or
MiniBooNE~\cite{MiniBooNE:2017nqe,MiniBooNEDM:2018cxm} experiments. Upper bounds can be derived
using narrow-width approximation by taking into account existing bounds for dark photon mixing
for mass $m_{A'}\sim m_\mu$ and without taking into account
intermediate vector meson states. In the hadrophilic case, we can obtain limits
for decay branching $\mathrm{Br}(\eta \to \gamma \mu e) <  0.08  \times (\alpha_B /\alpha)$
and $\mathrm{Br}(\eta^\prime \to \gamma \mu e) <  0.76 \times  (\alpha_B/\alpha )$,
where $\alpha_B=g_{qq}^2/4\pi$. From the COHERENT experiment one gets
$\alpha_B \sim 10^{-7} \div 10^{-9}$ for $m_\mu<m_A<m_P$. Finally, using the bound for the hadrophilic scenario we obtain
$\mathrm{Br}(\eta \to \gamma \mu e) \lesssim 2.7 \times 10^{-7} $ and
$\mathrm{Br}(\eta^\prime \to \gamma \mu e) \lesssim  2 \times 10^{-8}$.
For general case of dark vector boson scenario, we can use bounds from
the BABAR~\cite{BaBar:2017tiz} and NA64~\cite{NA64:2023wbi} experiments,
$\lesssim 2 \times 10^{-7}$ and $\lesssim 10^{-9}$, respectively.
Note, the branchings for the leptophilic case are suppressed by a factor $\alpha/64\pi^3$
in comparison with the general dark photon scenario. 
If dark photon mass $m_{A'}$ dark photon is heavier than mass of decaying meson,
then the branching is more suppressed due to existing bounds from LFV muon decays. 
The contribution with taking into account that intermediate meson states produce less
contribution are  omitted here, as they do not provide  new strict upper bounds.

For the case when diagonal coupling is dominant, $\Gamma(P \to \gamma \mu e)$ 
is proportional to $m_P \times g_{e\mu}^2$. 
For the general case of dark vector boson scenario, we can use bounds from 
and NA64~\cite{Zhevlakov:2023jzt} experiments for the LFV case,
$\lesssim 2 \times 10^{-12}$ and $\lesssim 10^{-13}$, respectively, for 
$\eta$ and $\eta^\prime$ mesons decays.

\section{Summary}
\label{Sec:summary}

We discussed a possibility to study $e-\mu$ conversion on nuclear
in experiments on fixed target in a phenomenological approach with
vector boson mediator (dark photon). 
Obtained results solidify the impossibility of such study due to the requirement
to increase statistics by orders of magnitude. The strict bounds from LFV muon decays.
Branchings of the $\eta$ and $\eta^\prime$ meson decays will be suppressed
when the vector meson mediator is heavier than $\eta$ and $\eta^\prime$ states.
In the resonance case, obtained  upper bounds will be
$\mathrm{Br}(\eta \to \gamma \mu e) < 2 \times 10^{-7}$
and $\mathrm{Br}(\eta^\prime \to \gamma \mu e) < 2\times 10^{-8}$ based on data limits 
collected from search for the dark photon in different experiments. This case also
looks problematic for study at future $\eta$ and $\eta^\prime$ factories but possible 
with planned meson yield~\cite{Chen:2024wad},
if the corresponding processes involve feebly interacting vector mediator 
(\textit{\`{a} la} dark photon). Additionally, decay into $\eta \to \gamma \mu e$ has a specific signature 
(with one muon in final) which can be detected and should be analyzed in super-$\eta$ factories.     
For the case of dominance of diagonal coupling the bounds will be more stringent 
at level $\lesssim 2 \times 10^{-12}$ and $\lesssim 10^{-13}$ for $\eta$ and $\eta^\prime$ meson decays, respectively. It will require high statistics of meson yield. For the case of dominance of dark mode, study of such rare decays will be unattainable. 

\begin{acknowledgments} 
	
	We would like to thank Dmitry Kirpichnikov and Xurong Chen for discussions.  
	This work was funded by FONDECYT (Chile) under Grant 
	No. 1240066 and by ANID$-$Millen\-nium Program$-$ICN2019\_044 (Chile).
	The work of A.~S.~Z. is supported by the Foundation 
	for the Advancement of Theoretical Physics and Mathematics "BASIS"
	and by the PIFI CAS Grant No.:2024PVB0070. 
	A.~S.~Zh. also is grateful Xurong Chen (IMP CAS, China)
	for the warm hospitality in Southern Center for Nuclear Science Theory,
        Institute of Modern Physics in Huizhou, where part of work was done. 
	
\end{acknowledgments}


\appendix
\onecolumngrid
\onecolumngrid
\section{Form factors describing interaction of external electromagnetic field with LFV current}
\label{Sec:AppA}

\eq
F^{lf}_2(q^2)&=&
\int\limits_0^1 \limits dx dy dz \delta(1-x-y-z)\frac{(m_f+m_l)x}{8\pi^2}
\Bigg(\biggl [ \frac{m_l(1+y)+m_f(z-1)}{D(m_l,m_f,y,z,t)}\biggl]
+ [m_l\leftrightarrow m_f,y\leftrightarrow z]\Bigg)\,,
\en
with $D(m_l,m_f,y,z,t)=m_{A^\prime}^2 x-z(m_f^2 x+ty)+m_l^2(y^2+z+yz)$.

\eq
F^{lf}_{1;r}(q^2)&=&\int\limits_0^1 \limits dx dy dz \delta(1-x-y-z)
\frac{1}{8\pi^2}\Bigg( \Biggl
     [\frac{2(m_l  m_f x (2-x) + m_l^2 (x - y - z))}{D(m_l,m_f,y,z,0)}
\\
&-&\frac{2 ( m_l^2 (x - y - z) + m_f m_l x (2-x) + t (x- y z) )}{ D(m_l,m_f,y,z,t)}
+ \ln \biggl( \frac{D(m_l,m_f,y,z,0)}{D(m_l,m_f,y,z,t)} \biggl) \biggl ]
+ [m_l\leftrightarrow m_f,y\leftrightarrow z]\Bigg) \,.\nn
\en

\twocolumngrid

\end{document}